\documentclass[journal]{IEEEtran}
\usepackage{blindtext}
\usepackage{graphicx}
\usepackage{amsmath}
\usepackage{amssymb}
\usepackage{listings}
\usepackage{xcolor}
\usepackage{cite}
\usepackage{hyperref}

\begin{document}

\title{Set-Based Adaptive Safety Control}

\author{Prithvi Akella, Sean Anderson, David Lovell}

\maketitle

\begin{abstract}
Feedback Control Systems, ME C134/EE C128, is an introductory control systems course at UC Berkeley.  Over the entire course, students gain practical experience by implementing various control schemes and designing observers in an effort to ultimately stabilize an inverted pendulum on a linear track.  Throughout this learning process, frequent mishaps occur where improper controller implementation damages hardware.  A simple example concerns the student's controller driving the cart into the wall at full speed.  To offset the financial burden placed on the university in light of these mishaps, we designed a streamlined adaptive control system using set theory.  We utilized lab-provided plant models to generate an $O_\infty$ set, attenuated the vertices to generate a safe, sub-region $S_\infty$, and attenuated in such a manner as to ensure an evolution of the vertices of $S_\infty$ remained within $O_\infty$ for at least one time step.  Afterwards, we constructed a single Simulink block for students to easily implement within their own control schemes.  This block consistently checks to see whether the system state remains within $S_\infty$.  If that check is true, our controller does nothing.  If it returns false, our controller takes over, drives the system to a prescribed safe-point, and shuts the system down.  Overall, our process assumes perfect plant modelling, though our insistence on an evolution of $S_\infty$ remaining within $O_\infty$ resulted in considerable robustness to disturbances.  In the end we were successful in implementing this real-time adaptive system and will provide it to the department for use in future labs.\\
Video Link: \url{https://drive.google.com/file/d/1kPw5FTVs3hOcy_yhZ8gXrCNN5lqBWi0v/view?usp=sharing}
\end{abstract}

\vspace{-0.15in}
\section{Introduction}
\vspace{-0.2in}
In an introductory controls class at Berkeley, students attempt to stabilize an inverted pendulum on a linear track \cite{lab_pend}.  Often, students implement unstable or otherwise flawed controllers that cause a variety of issues. This can cause the cart to drive into one of the side-walls subsequently destroying hardware, or cause the system to act wildly posing a risk to the students. Damage to this commercial lab equipment represents a non-trivial financial cost to the university, and injuries to students are unacceptable. To mitigate these issues, we created an adaptive control scheme that monitors and overrides the student's controller in the event of unsafe actions. In that effort, we note that the action of the cart running into the wall can be extrapolated from system dynamics. That is to say, given a specific braking scheme, we can use set theory to identify all possible states for which the aforementioned scheme can feasibly prevent unsafe actions. Since the actuation of the system occurs from a single motor voltage input that directly controls the position of the cart on the track, it sufficient to only check if the cart position leaves a region defined as safe.  Provided that region is a subset of the larger, calculated region, our controller will still function to bring the system back to a predefined safe point.

\section{Background Information}

\subsection{Stable Controller Development}
To clarify, $\mathcal{O}_{\infty}$ is a positive invariant set  defined as follows \cite{borrelli}:
\begin{align*}
    & \forall x(0) \in O \quad x(t) \in O \\ & \forall t \in [0,\infty] \quad | \quad x(k+1) = f(x(k)).
    \label{O_infty}
\end{align*}
Note that $f(...)$ in the above equation represents the linearized evolutionary scheme we developed by Euler-discretization of a lab-developed transfer function.  More specifically, for 
\begin{align*}
    & y^{(n)} + \alpha_1y^{(n-1)} + ... = u^{(p)} + \beta_1u^{(p-1)} + ..., \\
    & y_{k+1} = y_k + \Delta t \Dot{y}(k) \quad \Dot{y}_{k+1} = \Dot{y}_k + \Delta t \Ddot{y}(k) \quad ....
\end{align*}
The successive Euler-discretization steps can be reformatted into a matrix equation as
\begin{equation*}
    y_{k+1} = Ay_k + Bu_k = f(x_k,u_k).
    \label{linear_evolution}
\end{equation*}
Note that the $y_{k+1}$ in the above equation corresponds to a vector whose elements are $y_k$, $\Dot{y}_k$, and so forth.  However, O's functional dependence is only restricted to functions of x.  To remedy that, we develop a specific, stable controller such that each $u_k = -K x_k$.  Stability in the discrete scenario implies that the closed-loop evolution, $A-BK$ decays to $0$ after multiple iterations.  That is,
\begin{equation*}
    (A-BK)^N \rightarrow 0 \quad \textrm{as} \quad N \rightarrow \infty.
\end{equation*}
It is also critical to mention that the aforementioned stability implies that the poles of the Closed Loop System, $A_{cl} = A-BK$ all lie within the unit circle on the imaginary plane. $MATLAB$ employs that last criteria to generate stable controllers through the place function as shown below:
\begin{lstlisting}
    K = place(A,B,[p_1,p_2,...]).
\end{lstlisting}
Note that each of the $p_i$ correspond to the a pole location for the closed-loop system.  Choosing each $|p_i| < 1$ identifies a stable controller as per our definitions earlier.

\subsection{Calculation of $\mathcal{O}_{\infty}$}
$\mathcal{O}_{\infty}$ is found through a recursive process detailed below:
\begin{lstlisting}
    O = X;
    numiterations = 500;
    for i = 1:numiterations
        Pre_O = Pre(O);
        if Pre_O.intersect(O) == O
            Oinf = O;
            break;
        else
            O = Pre_O.intersect(O);
        end
    end
\end{lstlisting}
To clarify, the initial line setting $\mathcal{O}=\mathcal{X}$ concerns initializing the invariant set to be the initial, invariant, state-constraint set, $\mathcal{X}$. As a result, intersecting the set each iteration ensures that the resulting $\mathcal{O}_{\infty} \in \mathcal{X}$.  Furthermore, the $Pre$ operation is defined as follows:

\begin{equation*}
    Pre(G) = S = \{x \, | \, y = f(x) \, \forall y \in G \}.
\end{equation*}
Note that $\mathcal{G}$ is some predefined set, and for our purposes, $f(...)$ is the closed-loop evolutionary scheme we developed with $K$. The number of iterations is limited to prevent against infinite loop calculations. It is important to qualify that both of the aforementioned processes, generating $K$ and calculating $\mathcal{O}_{\infty}$ are model-specific processes. As such, this process as stated, assumes a perfect model with no noise or disturbances. To protect our process against faults in those assumptions, we implement a set attenuation and boundary evolution process detailed in the following implementation section.

\section{Implementation}
\subsection{Set Generation}
For our specific system, our continuous time matrices were,
\[
    A = 
        \begin{bmatrix}
            0 & 1 \\
            0 & -7.2
        \end{bmatrix} \quad
    B = 
        \begin{bmatrix}
            0  \\
            1.6
        \end{bmatrix},
\]
which, given a sample time of 0.002 seconds, transformed to the following discrete time matrices:
\[
    A = 
        \begin{bmatrix}
            1 & \Delta t \\
            0 & 1-7.2\Delta t
        \end{bmatrix} \quad
    B = 
        \begin{bmatrix}
            0  \\
            1.6\Delta t 
        \end{bmatrix}.
\]
Afterwards, we placed our desired closed loop poles at $ \lambda_1 = 0.99$ and $ \lambda_2 = 0.985$, which generated the following K controller:
\[
    K = 
        \begin{bmatrix}
           23.3 & 3.3
        \end{bmatrix}
\]
In addition, our only state constraints concerned the position limitations on the cart, while the velocity could feasibly be any real-value.  In reality, the physical constraints, predefined evolution, and specific controller should limit velocity, but we had no constraints a priori. Noting that, our initial $\mathcal{X}$ set was defined as,
\begin{equation*}
    \mathcal{X} = \{ x \, | -0.4<=x(1)<= 0.4 \quad \forall \, x(2) \in \mathbb{R} \}.
\end{equation*}
After following the aforementioned process to create $\mathcal{O}_{\infty}$ we decided that the two main limitations inherent to our system could be addressed by implementing a buffer region. Namely, this system is only valid with respect to our specific plant model, and that we could not use the calculated $\mathcal{O}_{\infty}$ as a boundary for the prescribed safe zone.  The second limitation arises if you consider an event where the system state left $\mathcal{O}_\infty$.  In this event, no controller would be able to control the system back to a safe region, as the state is already outside its region of attraction.  To remedy both, the vertices of $\mathcal{O}_{\infty}$ were found and scaled down to create $\mathcal{S}_{\infty}$. This set represents the the set of states that if violated give adequate response time for the system to be prevented from exiting $\mathcal{O}_{\infty}$.

\begin{figure}[!h]
    \centering
    \includegraphics[width=\linewidth]{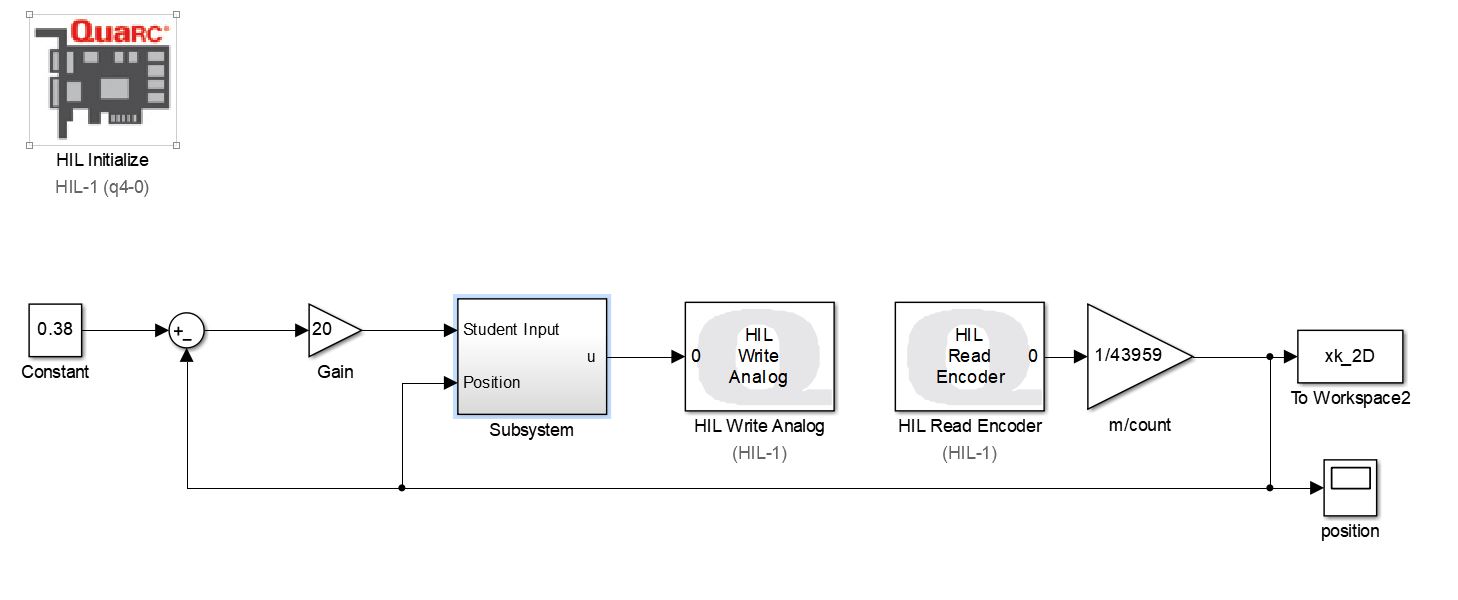}
    \caption{The background controller can easily be incorporated into the Simulink model by the student.}

    \label{controller}
\end{figure}

The entire system can be distributed to the students as a single Simulink sub-system block which is placed in the feedback loop between the student's controller and the input to the plant. The simplicity of this application is shown in Figure \ref{controller}. The inputs and outputs to the subsystem are labeled for ease of use.

\begin{figure}[!h]
    \centering
    \includegraphics[width=\linewidth]{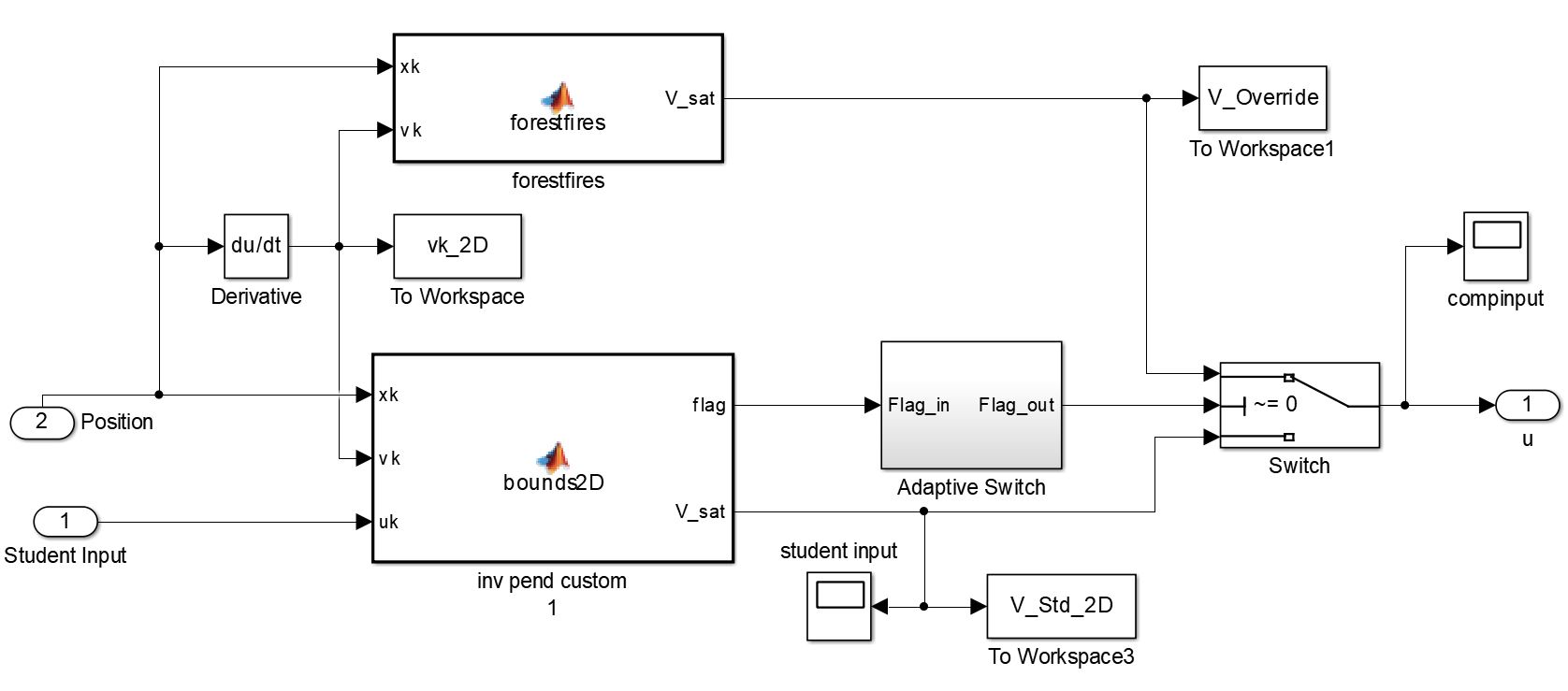}
    \caption{The subsystem implementing the background controller mainly consists of two user-defined functions.}
    \label{subsystem}
\end{figure}

The contents of the sub-system is shown in Figure \ref{subsystem}. It contains two MATLAB function blocks, a variable switch, and a switch logic block. The function, $bound2D$, which can be found in the appendix, implements the real time monitoring of the system and produces a flag if the system will violate constraints. If a flag is thrown, the switch logic prevents the student's input from feeding to the motor and instead feeds the signal through the $forestfires$ function block to drive the system to the defined zero point. The adaptive switch subsystem counts the number of flags and terminates the Simulink model at a defined time. If there is no flag, the student's input is passed through after being saturated to prevent damage to the physical system from large inputs.

\subsection{Robustness Measures}
To ensure that our controller would always bring the cart back to a safe region and that our system was relatively robust to model errors, noise, and disturbances, we attenuated $\mathcal{O}_{\infty}$ by scaling the vertices of the set by a scalar less than 1.  In doing so, we generated a subset of $\mathcal{O}_{\infty}$, $\mathcal{S}_{\infty}$ while ensuring that the evolution of every vertex of $\mathcal{S}_{\infty}$, with respect to the worst possible input for that state, remained within $\mathcal{O}_{\infty}$. This method created a "gray-zone" that is the difference between the sets. This gray-zone represents combinations of states that are outside of the predefined "safe" region, but still remain controllable. The process that was used to develop the gray-zone, evolving the boundary with the worst case input, provided the largest possible buffer region that simultaneously did not overly limit the working space along the track. If the working space on the track was too severely limited, it would effect the student's ability to conduct lab exercises. This gray-zone hedges against disturbances and modelling errors insofar as it provides the largest area to "catch" errors of the kind that would cause a danger to the lab equipment or students.  In addition, this process resulted in a system so robust that when the result of the 2D (cart-only) system was tested on the 4D (cart-and-pendulum) system, it proved to be sufficient.

\section{Results}
\begin{figure}[!h]
    \centering
    \includegraphics[width=\linewidth]{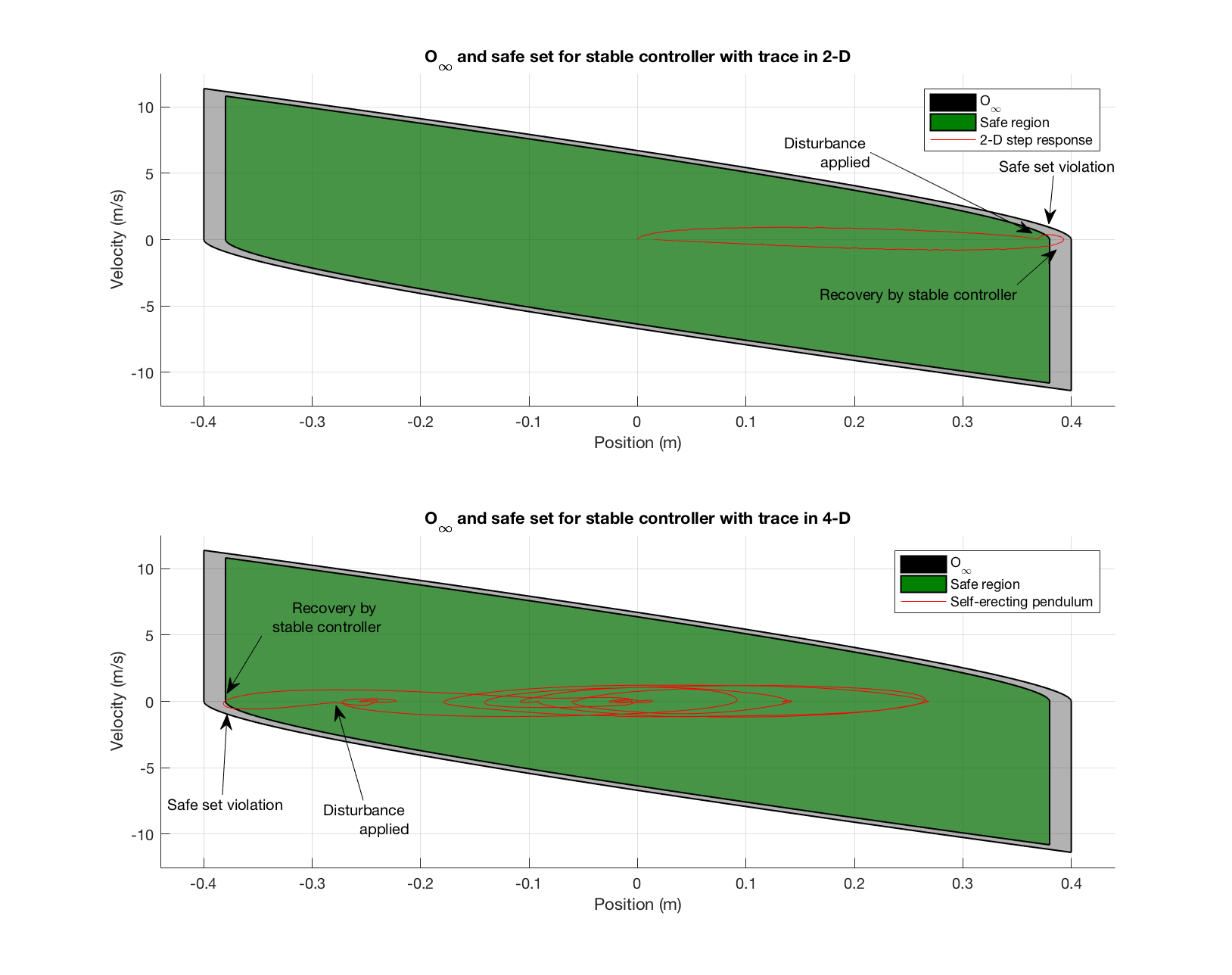}
    \caption{The upper image depicts $\mathcal{O}_{\infty}$, $S_{\infty}$, and the trace of one instance of the 2-D system with $S_{\infty}$ violation. The lower image uses the same set but the trace is for the cart-pendulum self-erecting scenario. }
    \label{result}
\end{figure}

The outlined process worked across all testing scenarios without failure. $\mathcal{S}_{\infty}$ sets were calculated for both the 2D and 4D systems, however, the 4D process restricted the usable track length too severely. Rather than construct a different method for the monitoring of the 4D system, the 2D system was tested to determine its effectiveness. These tests proved that the 2D system was fully capable of monitoring and overriding the 4D system. A characteristic test result of the 2D system acting on both systems are shown in Figure \ref{result}.

In Figure \ref{result}, $S_{\infty}$ is identified as the green area, with $\mathcal{O}_{\infty}$ as the larger gray area. The difference in areas, the 'grey-zone', is the set of states that violate the constraints of $\mathcal{S}_{\infty}$, but remain within  $\mathcal{O}_{\infty}$. This 'grey-zone' gives the override controller sufficient time to physically stop the system from violating $\mathcal{O}_{\infty}$ by driving the cart away from the wall. In the top image of Figure \ref{result}, the 2D system is shown. The red line indicates the state trajectory of the cart during a step input which places it at the boundary of $\mathcal{S}_{\infty}$. At this point, a disturbance is applied to the cart which causes the system to leave $\mathcal{S}_{\infty}$. The monitoring system identifies this violation and initiates the override controller to bring the cart back to the defined zero point. Note that this proves an instance of our controller's robustness to disturbances. Even though the disturbance that caused the violation was not factored into our set calculation, our controller still acted properly in the event of the failure and brought the system to safety.

Additionally, the lower image of Figure \ref{result}, exhibits a characteristic response of the 2D monitoring and override system acting on the cart with the pendulum added. In this scenario the pendulum started at rest in the vertically down position. The axes of this plot are the position and velocity of the cart, as such, the trajectory of states for the pendulum are not shown. As the student's control system initiates a pumping sequence, the position and velocity of the cart oscillate from positive to negative until the pendulum is erected and stabilized near the left boundary of the track. This process verifies that $\mathcal{S}_{\infty}$ does not overly restrict our available region. That is, there is more than sufficient track length for the pendulum to be erected and balanced. 

After the cart stabilized the pendulum, a disturbance was applied to the pendulum. The cart then acted to attempt to regain balance of the pendulum and in the process the $\mathcal{S}_{\infty}$ set was violated. Again, it is interesting to note that the disturbance to the pendulum was responsible for the violation of $\mathcal{S}_{\infty}$, but was not considered in the calculation. Still, the override system responded correctly to prevent the cart from colliding with the wall. In this way, the addition of the pendulum to the 2D system can be interpreted simply as a disturbance. This disturbance can cause the cart to violate $\mathcal{S}_{\infty}$, which the monitoring system then identifies causing the override system to respond. Also of note is that the unpredictable inertia of the swinging pendulum does not prevent the override system from recovering the cart before a violation of $\mathcal{O}_{\infty}$ occurs. This further proves the robustness of our system. 

\section{Conclusion}
We successfully created a simple Simulink block that monitors and overrides a student's controller in the event of a hardware failure or safety concern. The application of set theory provided a means to accurately determine the safe operating length of the track. The resulting matrices allowed for real-time identification of state violations even with a controller operating in millisecond sample periods. This system also proved to be robust to the extent that a high degree of modifications to the mass and inertia of the plant were successfully tolerated.

\newpage
\section{Appendices}
\subsection{Main Code}
\begin{lstlisting}
% Kd = 8.56;
% Kp = 65.6;
% K = [Kp Kd];

clc
close all
dt = .002;
Adt = [1 dt; 0 1-7.197*dt];
Bdt = [0; 1.606*dt];
K = place(Adt,Bdt,[0.99 0.985]);
scale = .95;
disp(K);
Acl = Adt - Bdt*K;

nu = size(Bdt,2);
xl = [-.4, -Inf];
xu = [.4, Inf];
ul = -6;
uu = 6;

% find Oinf for discrete controlled system
X = Polyhedron('lb',xl,'ub',xu);
U = Polyhedron('lb',ul,'ub',uu);
S = X.intersect(Polyhedron('H',[-U.H(:,1:nu)*K U.H(:,nu+1)]));
S = X;
Oinf = max_pos_inv(Acl,S);

% plot Oinf and a scaled version (based on vertices)
figure;
plot(Oinf,'color','red','alpha',.5); hold on;
plot(S,'color','blue','alpha',0.5);
vertices = Oinf.V;
numVert = Oinf.minHRep();
scaled_vertices = vertices*scale;
figure(2); hold on;
O2 = Polyhedron('V',scaled_vertices);
plot(Oinf,'color','black','alpha',0.3);
plot(O2,'color','green','alpha',.6);
xlabel('Position (m)'); ylabel('Velocity (m/s)');
legend('O_{\infty}','Safe region');
title('O_{\infty} and safe set for stable controller');


% step system forward one step with worst case scenario
% worst case scenario is taken to be full acceleration
pairs = size(scaled_vertices,1);
H = Oinf.H;
Aconst = H(:,1:end-1);
Bconst = H(:,end);
count = 0;
for i = 1:pairs
    x_now = scaled_vertices(i,:)';
    x_dir = sign(x_now(1));
    x_next = Adt*x_now + Bdt*(x_dir*6);
    FLAG = all(Aconst*x_next <= Bconst);
    count = count + double(FLAG);
end
disp(count == pairs);
s = O2.H;
Aconstr = s(:,1:end-1);
Bconstr = s(:,end);
save('allowable_states_4d.mat','Aconstr','Bconstr','Oinf','O2');

\end{lstlisting}
\subsection{bounds2D}
\begin{lstlisting}
function [flag,V_sat]  = bounds2D(xk,vk,...
    uk, Aconst, Bconst)
% Goals:
% a) Compute the evolution of the states based on intended input
% b) Verify that I don't violate my state constraints
% c) if I don't... do absolutely nothing
% d) If I do, flag the system to stop

%%%%%%%%%%%%%
% Saturate input so that the model predicts accurately
%%%%%%%%%%%%%
max_lim = min(6,max(4,10*vk+5));
min_lim = min(-4,max(-6,10*vk-5));
% Dynamic Saturations
V_sat = max(min_lim,uk);
V_sat = min(max_lim,V_sat);

%%%%%%%%%%%%%
% Create state vector for S_inf violation calculation
%%%%%%%%%%%%%
x_now = [xk; vk];

%%%%%%%%%%%%%
% Check to see if a violation will occur
%%%%%%%%%%%%%
flag = ~all(Aconst*x_now<= Bconst);
% Note that if the above flag is true, that means the evolved state will
% violate constraints. If no violation occurs flag = 0.
end
\end{lstlisting}
\subsection{forestfires}
\begin{lstlisting}
function V_sat = forestfires(xk,vk)
% This code acts as an override controller to drive the cart away from the
% wall to the defined zero point.

%%%%%
% set responsive gains for fast action
%%%%%
K = [23.3499 3.3020];

%%%%%
% identify state vector
%%%%%
statek = [xk;vk];

%%%%%
% state control law
%%%%%
uk = -K*statek;

%%%%%
% saturate input to prevent damage to physical system 
%%%%%
max_lim = min(6,max(4,10*vk+5));
min_lim = min(-4,max(-6,10*vk-5));
% Dynamic Saturations
V_sat = max(min_lim,uk);
V_sat = min(max_lim,V_sat);
end
\end{lstlisting}

\section*{Acknowledgment}
$Pre$ function and set theory were sourced from ME C231A-EECS C220B, UC Berkeley, Fall 2017 course material.

System dynamics, gains, and state matrices for the lab hardware were sourced from ME C134-EECS C128, UC Berkeley, Fall 2017 course material.


\begin{thebibliography}{19}
\bibitem{lab_pend} 
Staff ME C134/ EECS 128. "Lab  6d:  Self-erecting  inverted  pendulum  (seip),  mec134/ eecs c128,” April 2017.

\bibitem{borrelli}
F. Borrelli,  “Lecture  notes:  Model  predictive  control  reachability  and invariance, eecs c220b/me c231a,” November 2017.

\end{thebibliography}
\end{document}